# Quantum Carry-Save Arithmetic

Phil Gossett
**Silicon Graphics, Inc. 2011 N. Shoreline Blvd. Mountain View, CA 94043-1389**
e-mail:pg@engr.sgi.com
**August 29, 1998**

**Abstract: This paper shows how to design efficient arithmetic elements out of quantum gates using "carry-save" techniques borrowed from classical computer design. This allows bit-parallel evaluation of all the arithmetic elements required for Shor's algorithm, including modular arithmetic, deferring all carry propagation until the end of the entire computation. This reduces the quantum gate delay from $O(N^3)$ to $O(N \log N)$ at a cost of increasing the number of qubits required from $O(N)$ to $O(N^2)$.**

## 1.0 Introduction

Of the recent advances in quantum algorithms, one of the most impressive to date is Shor's algorithm for discrete logs and factorization [1], which gives an exponential speedup over classical algorithms. Vedral, Barenco and Ekert [2] have shown how to implement the necessary modular exponentiation operations in quantum gates with a number of qubits linear in the number of input bits. These networks use "ripple carry" for the propagation of carries in the additions required for the computation. Ripple carry adders have delay linear in the number of bits.

Classical computer designs have long used "carry-save" addition [3] to defer carry propagation until the very end of a computation, with the intermediate additions having delay constant in number of bits. Since these can be arranged in binary trees, the cumulative gate delay for a classical carry-save implementation is logarithmic in number of bits.

This paper shows how to implement carry-save arithmetic with quantum gates. While requiring polynomially more qubits, we can retain the classical logarithmic delay, while still taking advantage of quantum superposition. For Shor's algorithm, modular arithmetic is needed. The implementation of [2] accomplishes this with operations conditional on comparisons of intermediate results. These comparisons amount to carry propagation. This paper also shows how modular arithmetic can be accomplished without carry propagation. Since Shor's algorithm is only interesting for inputs of at least hundreds of bits, this can be a significant reduction in gate delay.

## 2.0 Classical Carry-Save Addition

Classical adders (ripple carry or carry-save) are typically made out of "full adders", which are 3-input, 2-output devices. The classical full adder has the following truth table:

**TABLE 1. Classical Full Adder.**

```
Input      Output
A B Cin    Sum Cout
0 0 0      0   0
0 0 1      1   0
0 1 0      1   0
0 1 1      0   1
1 0 0      1   0
1 0 1      0   1
1 1 0      0   1
1 1 1      1   1
```

Note that the inputs are interchangeable, including the carry input (Cin). The Sum is the exclusive-or of the three inputs, while the carry output (Cout) is the "majority" of the same three inputs.

Multiple instances of full adders can be connected, carry out to carry in, to form a "ripple carry" adder:

**FIGURE 1. Classical Ripple Carry Adder.**

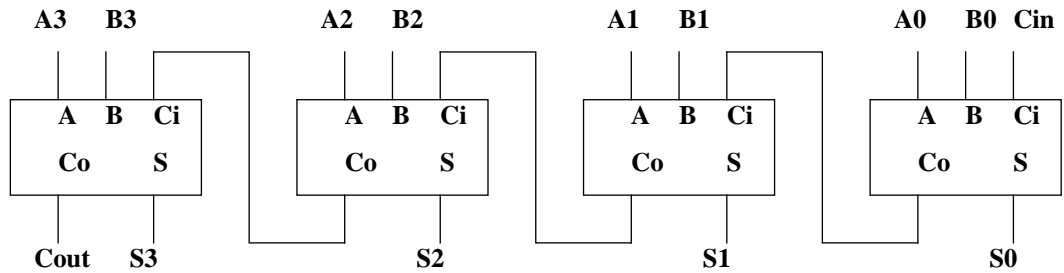

But since the inputs are interchangeable, we can use a stage of full adders as a 3-input, 2-output device, which adds three numbers, producing a sum and a carry output. This is called a "carry-save" adder.

**FIGURE 2. Classical 3->2 Carry-Save Adder.**

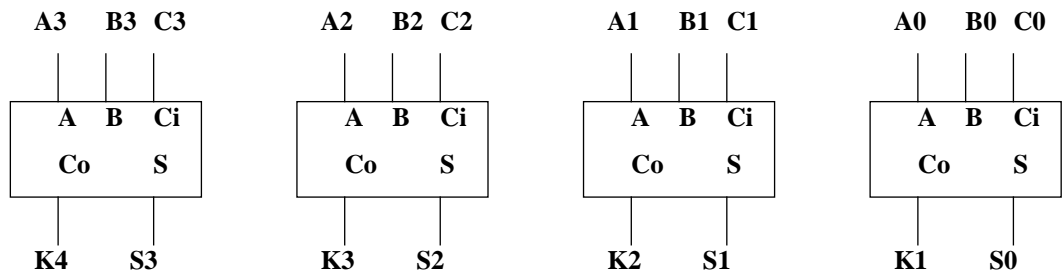

Adding the sum and carry outputs will produce the normal binary sum of the three original inputs, but this will require a carry propagation. If we chose not to propagate the carry, we can then add a fourth input to the two outputs from the previous stage of full adders, again producing a sum and a carry output.

**FIGURE 3. Classical 4->2 Carry-Save Adder.**

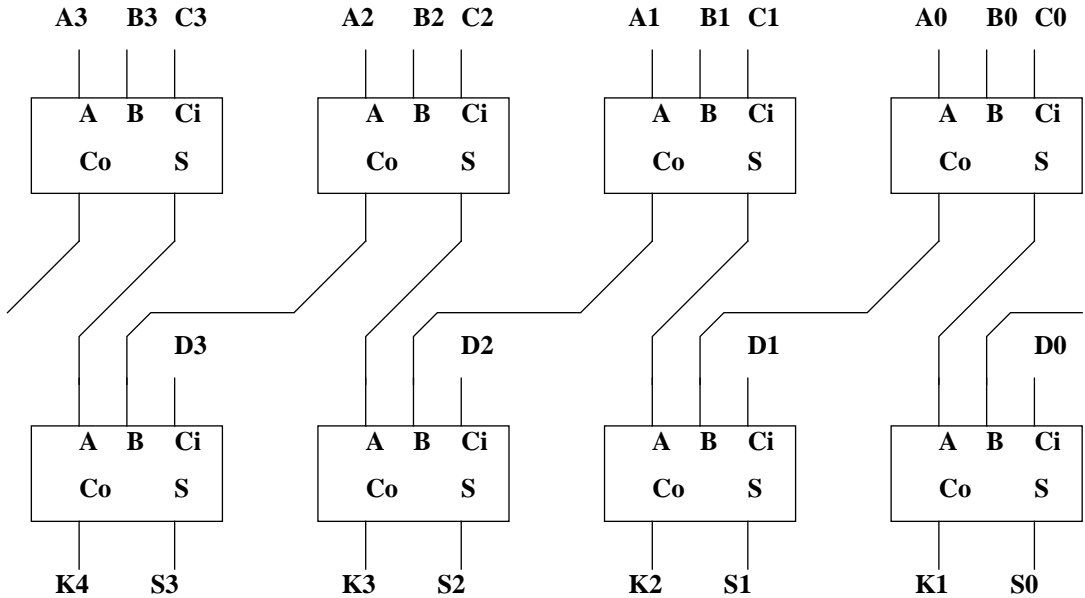

This process can be continued indefinitely, adding an input for each stage of full adders, without any intermediate carry propagation. These stages can be arranged in a binary tree structure, with cumulative delay logarithmic in the number of inputs to be added, and invariant of the number of bits per input.

## 3.0  Quantum Carry-Save Addition

Obviously, we can't directly implement carry-save adders with quantum gates, since the classical version of this element is clearly not unitary. It doesn't have as many outputs as inputs, so it can't be reversible. It's easy to see that even adding a third output isn't enough to make the full adder reversible. The truth table for the classical full adder (see above) has 3 inputs which map to a sum and carry of 1 and 0 respectively, and 3 inputs which map to a sum and carry of 0 and 1 respectively. Adding one bit obviously can't distinguish among 3 values. So we have to add two outputs, making the quantum equivalent of the classical full adder a 4-input, 4-output device.

The culprit with making the full adder unitary is the "majority" function for the carry. This is a fundamentally irreversible operation. However, with the addition of an ancillary input, any irreversible operation can be made reversible, since a reversible function of the ancillary input with any function (reversible or not) of passed-through inputs is reversible.

**FIGURE 4. Making an Irreversible Function Reversible.**

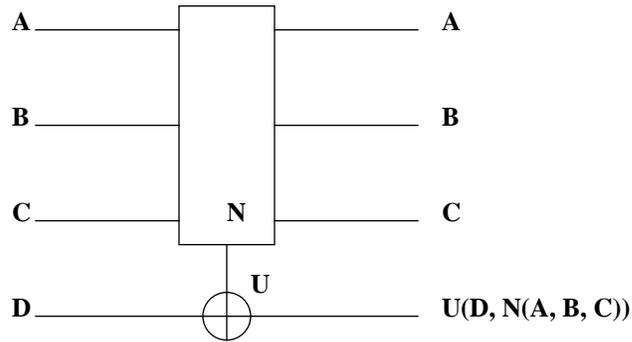

The simplest construction of a "quantum full adder" is to pass through two of the three classical inputs, and xor the new ancillary input with the majority function of the three classical inputs. This 4-input, 4-output device has the following truth table and equations:

**TABLE 2. Quantum Full Adder.**

```
Input      Output
D C B A    K S B A
0 0 0 0    0 0 0 0
0 0 0 1    0 1 0 1
0 0 1 0    0 1 1 0
0 0 1 1    1 0 1 1
0 1 0 0    0 1 0 0
0 1 0 1    1 0 0 1
0 1 1 0    1 0 1 0
0 1 1 1    1 1 1 1
1 0 0 0    1 0 0 0
1 0 0 1    1 1 0 1
1 0 1 0    1 1 1 0
1 0 1 1    0 0 1 1
1 1 0 0    1 1 0 0
1 1 0 1    0 0 0 1
1 1 1 0    0 0 1 0
1 1 1 1    0 1 1 1
```

$$S = Xor(A, B, C) \qquad (EQ\ 1)$$

$$K = Xor(D, Maj(A, B, C)) \qquad (EQ\ 2)$$

S is the sum output of the quantum full adder, and K is the carry out xored with the ancillary input D. This can be implemented with CNOT and Toffoli gates [4] as follows:

**FIGURE 5. Quantum Full Adder.**

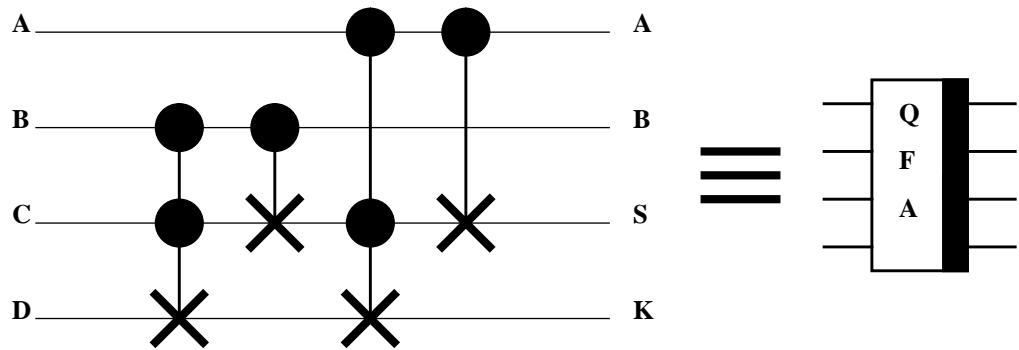

Note that this remarkably symmetrical implementation of a quantum full adder is just one additional CNOT gate appended to the "CARRY" gate of [2].

Using these quantum full adders, we can implement a ripple carry adder similar to the construction in [2]. But first, like that construction, we need a new gate to undo the carries, without disturbing the sums. This is a pure majority gate, without the sum part of the quantum full adder. This has the following truth table:

**TABLE 3. Quantum Majority Gate.**

```
Input       Output
D C B A     K C B A
0 0 0 0     0 0 0 0
0 0 0 1     0 0 0 1
0 0 1 0     0 0 1 0
0 0 1 1     1 0 1 1
0 1 0 0     0 1 0 0
0 1 0 1     1 1 0 1
0 1 1 0     1 1 1 0
0 1 1 1     1 1 1 1
1 0 0 0     1 0 0 0
1 0 0 1     1 0 0 1
1 0 1 0     1 0 1 0
1 0 1 1     0 0 1 1
1 1 0 0     1 1 0 0
1 1 0 1     0 1 0 1
1 1 1 0     0 1 1 0
1 1 1 1     0 1 1 1
```

Unlike the quantum full adder, we only implement the majority gate portion of the logic as above, with the K output given by EQ 2, but passing through the C input. This can be implemented with CNOT and Toffoli gates in a structure very similar to the quantum full adder, as follows:

**FIGURE 6. Quantum Majority Gate.**

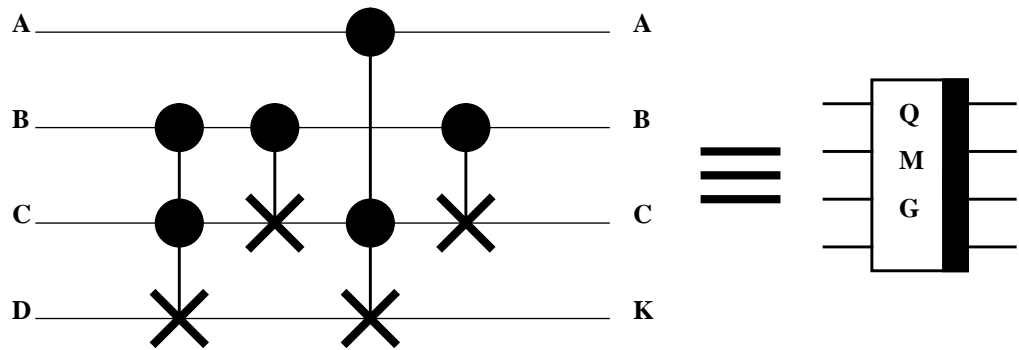

A network of quantum full adders and quantum majority gates can be constructed to implement a ripple carry adder similar to the one in [2] as follows:

**FIGURE 7. Quantum Ripple Carry Adder.**

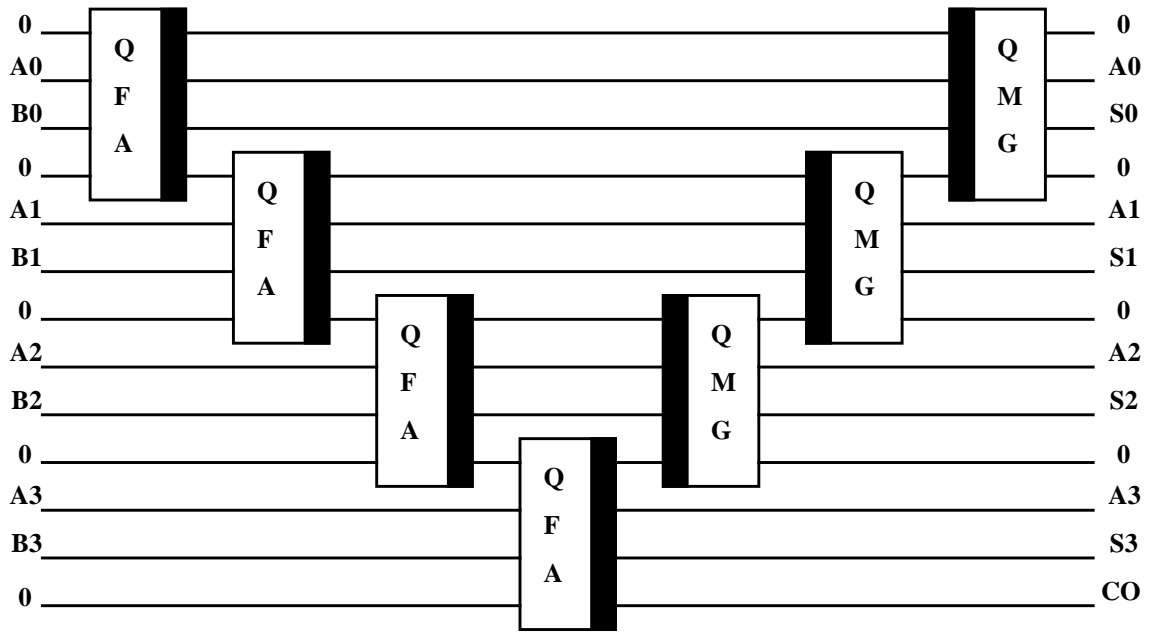

More interestingly, we can form a network for each bit of input that is the quantum equivalent of the classical 3->2 carry-save adder as follows:

**FIGURE 8. Quantum 3->2 Carry-Save Adder.**

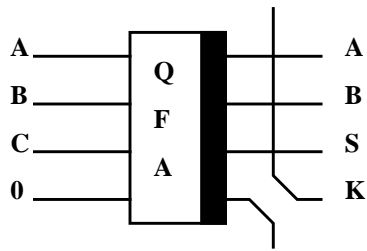

Note the carry out comes from the less significant bit above, and goes to the more significant bit below the one illustrated.

Extending this requires undoing any intermediate operations, so the quantum equivalent of the classical 4->2 carry-save adder is a bit more complicated:

**FIGURE 9. Quantum 4->2 Carry-Save Adder.**

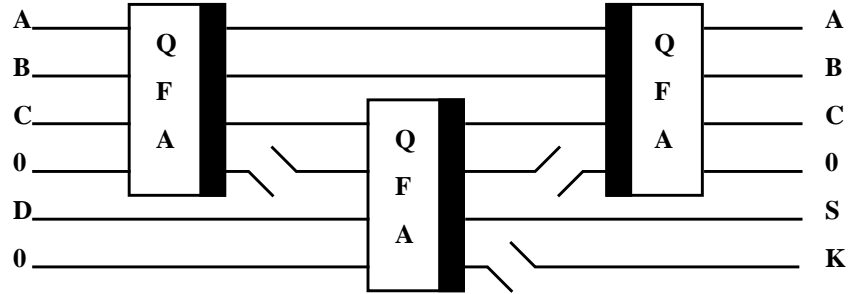

As in the classical implementation, this can be extended indefinitely in a tree structure. For example, an 8->2 carry-save adder can be implemented in a tree structure as follows:

**FIGURE 10. Quantum 8->2 Tree-Structured Carry-Save Adder.**

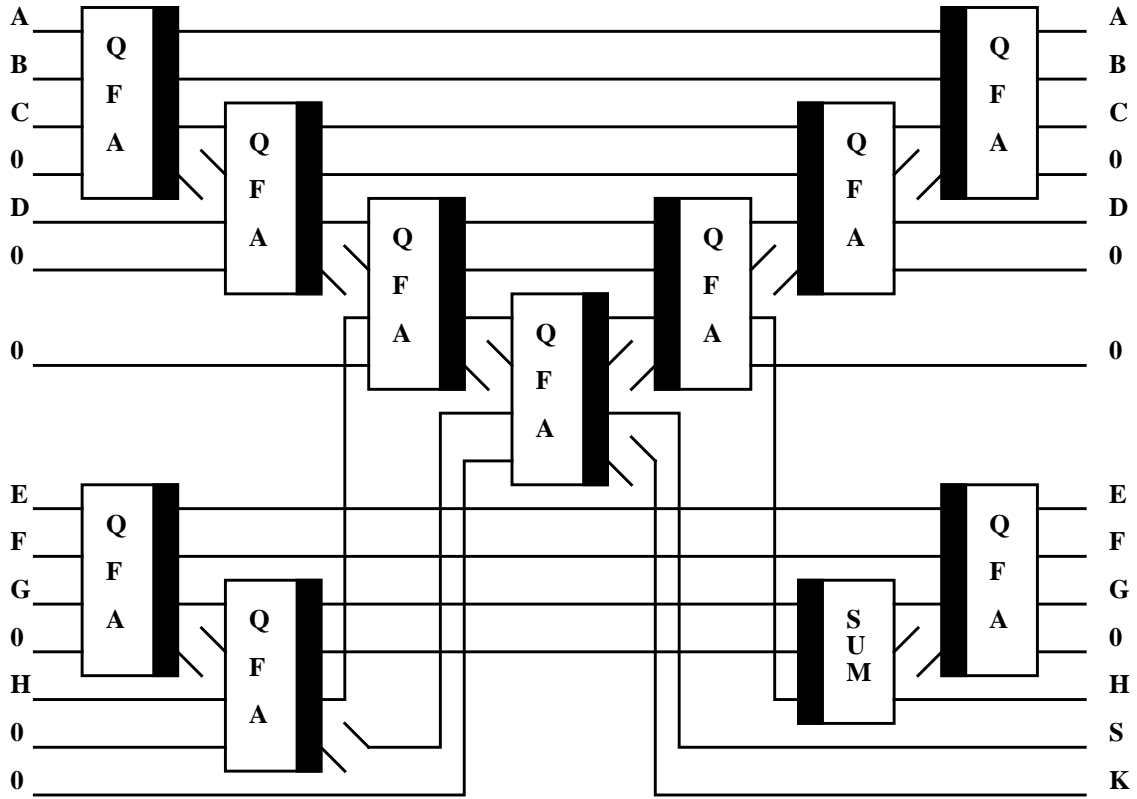

Here, the "SUM" gate is from [2], and is a 3-input xor with 2 inputs passed through. Note that for the general tree-structured carry-save adder, all the inputs must be passed through as outputs. The number of qubits required for N inputs of L bits each is (2N-2)L. The cumulative delay is $4\log_2(N)-5$ QFA delays, where the $\log_2(N)$ is rounded up to the nearest integer. Since in general, the inputs to the N-way adder are themselves in carry-save format, there are 2 binary inputs for each carry-save input. So, for N carry-save inputs, (4N-2)L qubits are required, with a cumulative delay of $4\log_2(N)-1$ QFA delays.

## 4.0 Quantum Carry-Save Multiplication

The carry-save implementation of a multiplier requires a number of qubits quadratic in number of input bits. However, since an N-bit multiply is the sum of N partial products, which can be done in a tree structure, the cumulative delay is only logarithmic in the number of input bits.

**FIGURE 11. Quantum 2-bit Carry-Save Multiplier.**

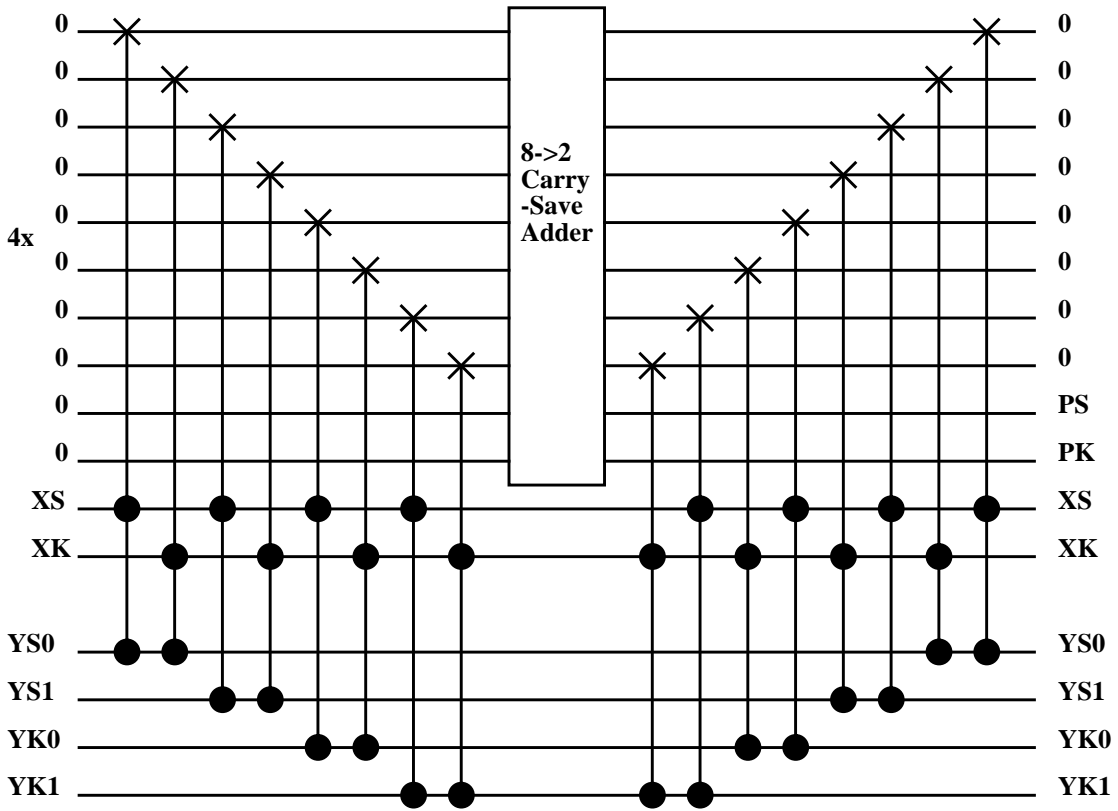

The partial product adder for an N-bit carry-save multiplier requires N partial products of 2N bits each, with a total of $8N^2-4N$ qubits, and a cumulative delay of $4\log_2(N)-1$ QFA delays. An additional 4N qubits are required to hold each of the two carry-save input pairs, for a total of $8N^2$ qubits, while an additional 2 Toffoli gate delays are required to create and destroy the partial products. Since two Toffoli gate delays are roughly equal to one QFA delay, the total cumulative delay is approximately $4\log_2(N)$ QFT delays.

## 5.0 Quantum Modular Carry-Save Arithmetic

If we were to use the method of [2] to implement modular arithmetic, we would be forced to propagate the carries at the end of each modular arithmetic operation, which would defeat the advantage of carry save arithmetic. Happily, this is not necessary. We can implement modular arithmetic by allowing the results to have some redundant representations. Note that we are already doing this with carry-save arithmetic, since there are multiple representations for any particular binary result. All we really need to insure (at least until the very end of the computation) is that there is no loss of information due to "wrapping" of intermediate results. This can be done by noting the following modular arithmetic identity:

$$x \equiv x \mathbin{\&} (2^N-1) + (x \mathbin{>>} N) * Q \quad \text{(modulo M)} \tag{EQ 3}$$

$$Q = 2^N \bmod M \tag{EQ 4}$$

where M is representable in N bits. Note that for Shor's algorithm, the modulus is a constant for the duration of the calculation. Hence, the calculation of Q (and P below) can be done "off-line" in the classical domain, for which there are well-known polynomial time algorithms [5].

It may not appear obvious that this actually limits the number of bits, since the addition could itself overflow. But it turns out that with the proper structure, and assuming we limit the modulus M to be no greater than $2^{N-2}$, we can insure that there will be no overflows. The following example should make this clearer:

**TABLE 4. Quantum 4-bit Modular 3->2 Carry-Save Adder "Worst Case" Example.**

```
  1111
  1111
 +1111
  P111
 QP11
 +  QQ    <= 11
    010     0 +
  0111    011  <= 3
 +  PP    <= 11
  1111
  0010    00
 +  PP    <= 11
   1000
  01110   0
```

$$P = 2^{N-1} \bmod M \tag{EQ 5}$$

$$Q = 2^N \bmod M \tag{EQ 6}$$

As long as the first set of italicized numbers (*0* + *011* in this example) is less than or equal to 3, which must be the case after the bits represented by **P** and **Q** are truncated, the most significant bit of the second set of italicized numbers (*00* in this example) will be zero, and the third single-bit italicized number (*0* in this example) will also be zero. This insures that there will be no overflows.

In order to construct an efficient network to implement quantum modular addition, we'll need the quantum equivalent of the classical "half adder". This is a 2-input, 2-output device with the following truth table:

**TABLE 5. Classical Half Adder.**

| Input | | Output | |
|---|---|---|---|
| A | B | Sum | Cout |
| 0 | 0 | 0 | 0 |
| 0 | 1 | 1 | 0 |
| 1 | 0 | 1 | 0 |
| 1 | 1 | 0 | 1 |

The sum output is the xor of the two inputs, and the carry output is the and of the same two inputs. The quantum equivalent is a 3-input, 3-output device with the following truth table and equations:

**TABLE 6. Quantum Half Adder.**

| Input | | | Output | | |
|---|---|---|---|---|---|
| C | B | A | K | S | A |
| 0 | 0 | 0 | 0 | 0 | 0 |
| 0 | 0 | 1 | 0 | 1 | 1 |
| 0 | 1 | 0 | 0 | 1 | 0 |
| 0 | 1 | 1 | 1 | 0 | 1 |
| 1 | 0 | 0 | 1 | 0 | 0 |
| 1 | 0 | 1 | 1 | 1 | 1 |
| 1 | 1 | 0 | 1 | 1 | 0 |
| 1 | 1 | 1 | 0 | 0 | 1 |

$$S = Xor(A, B) \qquad (EQ\ 7)$$

$$K = Xor(C, And(A, B)) \qquad (EQ\ 8)$$

S is the sum output of the quantum half adder, and K is the carry out xored with the ancillary input C. This can be implemented with one CNOT and one Toffoli gates as follows:

**FIGURE 12. Quantum Half Adder.**

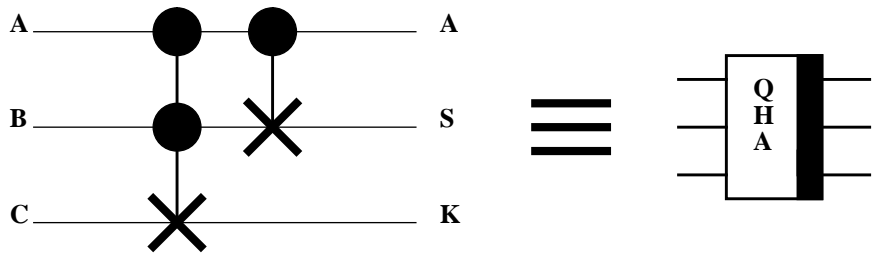

The following structure implements a quantum modular carry-save adder for the general case:

**FIGURE 13. Quantum Modular 3->2 Carry-Save Adder.**

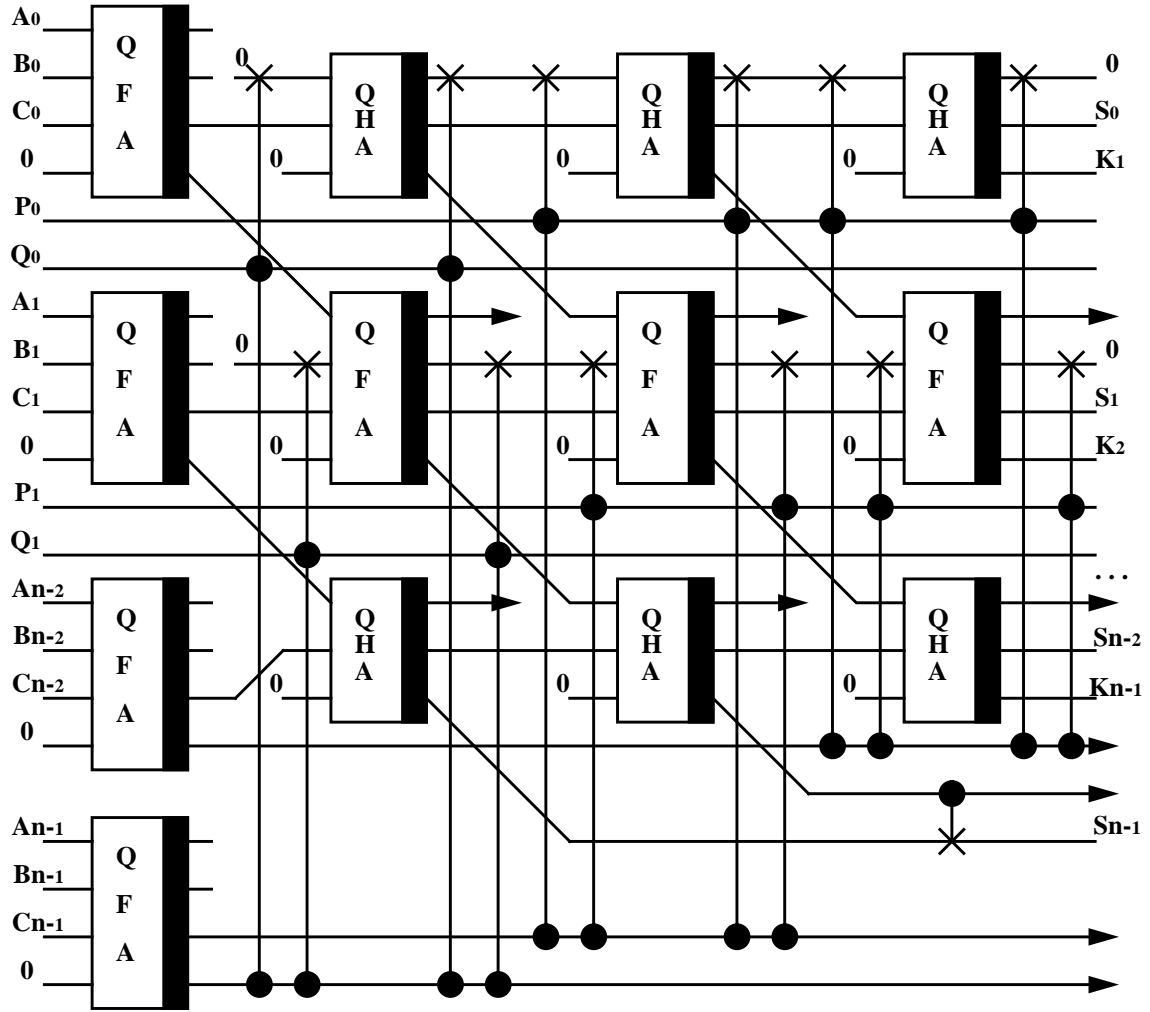

We can easily see that there can be no overflows, since the carry outputs from the two quantum half adders contributing to the $S_{n-1}$ output cannot both be one. The sum out of the first will be zero if its carry is one, and the carry from the second cannot be one if the first half-adder's sum is zero. Therefore, a CNOT gate can be used in place of a half adder to generate the most significant sum output bit, since the corresponding carry must always be zero.

Note that this will require more qubits of intermediate result, indicated above by arrows, which will have to be undone in a manner similar to that shown for multi-input adders and multipliers as described above. Given this, each 3->2 carry-save adder in the multiplier described above can be substituted with this structure of 4 sets of quantum full adders. This introduces a constant factor increase in delay and in qubits. But there is no carry propagation required.

# 6.0 Quantum Modular Carry-Save Exponentiation

Given modular multiplication, modular exponentiation can be implemented by a linear chain of multipliers, each either squaring the previous result or multiplying it by the base, according to each bit of the exponent. Since this is itself linear in multiplier delay, the net delay for a modular exponentiation is order N log N for carry-save arithmetic, compared with order $N^3$ for ripple carry arithmetic. The price we pay for this is that the number of qubits required is order $N^2$ for carry-save arithmetic, compared with order N for ripple carry arithmetic.

Note that as it is, the final result will still be in carry-save form. For some algorithms, this may be OK, leaving the final carry propagation and elimination of redundant modular representations to the classical domain. But for Shor's algorithm, it is necessary to produce a unique modular representation of the final result in the quantum domain. This can be done by summing the carry-save results with a ripple carry adder, and taking the true modulus by repeated trial subtractions. Since this only has to be done once at the very end of the computation, this doesn't significantly affect the total delay.

# 7.0 Conclusion

For N of 1000, a typical case for Shor's algorithm, there will be an effective speedup using carry-save versus ripple-carry arithmetic of a factor of roughly $(10^3)^3/(10*10^3) = 10^5$, at a cost of a factor of roughly $(10^3)^2/10^3 = 10^3$ qubits.

# 8.0 Acknowledgments

I would like to thank Christof Zalka for pointing out that the number of qubits required is $O(N^2)$, not $O(N^3)$ as I had mistakenly assumed in a previous version of this paper, and the need for converting to a unique modular representation in Shor's algorithm, as well as for many other helpful prior correspondences.